\documentclass[aps,prd,floatfix,preprint]{revtex4}

\usepackage{amsmath,amssymb,graphicx,color,braket}
\usepackage{hhline}

\newcommand{\dip}{Dipartimento di Fisica, Universit\`a di Trento\\
                                     via Sommarive 14, 38123 Trento, Italia\\}
\newcommand{\infn}{TIFPA (INFN)\\via Sommarive 14, 38123 Trento, Italia\ \medskip}
\newcommand{\maxR}[1]{Massimiliano Rinaldi${}^{#1}$\footnote{e-mail:\sl massimiliano.rinaldi@unitn.it\rm}}
\newcommand{\guido}[1]{Guido Cognola${}^{#1}$\footnote{e-mail:\sl cognola@science.unitn.it\rm}}

\newcommand{\luciano}[1]{Luciano Vanzo${}^{#1}$\footnote{e-mail:\sl vanzo@science.unitn.it\rm}}
\newcommand{\sergio}[1]{Sergio Zerbini${}^{#1}$\footnote{e-mail:\sl zerbini@science.unitn.it\rm}}

\newcommand{\be}{\begin{equation}}
\newcommand{\ee}{\end{equation}}
\newcommand{\bea}{\begin{eqnarray}}
\newcommand{\eea}{\end{eqnarray}}
\newcommand{\non}{\nonumber}

\begin{document}

\title{Inflation in scale-invariant theories of gravity}

\author{\maxR{}} \author{\guido{}}  \author{\luciano{}} \author{\sergio{}}
%\email{massimiliano.rinaldi@unitn.it}
\affiliation{\dip}
\affiliation{ \infn}

\date{\today}
%%%%%%%%%%%%%%%%%%%  ABSTRACT %%%%%%%%%%%%%%%

\begin{abstract} \noindent Thanks to the Planck Collaboration, we know the value of  the scalar spectral index of primordial fluctuations with unprecedented precision. In addition, the joint analysis of the data from Planck, BICEP2, and KEK has further constrained the value of the tensor-to-scalar ratio $r$ so that chaotic inflationary scenarios seem to be disfavoured. Inspired by these results, we look for a model that yields a  value of $r$ that is larger than the one predicted by the Starobinsky model but is still within the new constraints. We show that purely quadratic, renormalizable, and scale-invariant gravity, implemented by loop-corrections, satisfies these requirements.
\end{abstract}
 
 %%%%%%%%%%%%%%%%%%%%%%%%%%%%%%%%%%%%%%%%

\maketitle

%%%%%%%%%%%% INTRO  SECTION %%%%%%%%%%%%%%%%%%%%%

\section{Introduction}

%%%%%%%%%%%%%%%%%%%%%%%%%%%%%%%%%%%%%%%%

\noindent The claim of  the BICEP2 collaboration \cite{B14} that the tensor-to-scalar ratio has the surprisingly large value $r \simeq 0.2$, which points to a robust production of gravitational waves during inflation, has polarized the attention of the physics community. The situation has become less clear when some serious  criticisms to the BICEP2 analysis appeared in the literature (see for example \cite{Flauger}). Very recently, the joint analysis of data from Planck, KEK, and BICEP2 has finally settled the issue, fixing a new upper value for the tensor-to-scalar ratio at $r<0.09$  \cite{P15,B15}. In conclusion, although this bound  is much lower than initially claimed in \cite{B14}, we can say that it is still significant enough to allow for alternative theories to the Starobinsky Model (SM). 

As showed in \cite{staro}, cosmic inflation can be studied in models of gravity where the Lagrangian $\sqrt{g}R$ is replaced by a suitable function  $\sqrt{g}f(R)$ (see the general analysis in \cite{app}). Along these lines, in \cite{trento} we  showed that  the functional form of the inflationary $f(R)$ can be determined if the scalar spectral index $n_{s}$ and the tensor-to-scalar ratio $r$ are known with sufficient accuracy. In particular, we have shown that the slow-roll conditions imply that $f(R)\sim R^{2-\delta}	\sim R^{2}(1-\delta\log R+\cdots)$, where $\delta$ is a small and weakly time-dependent parameter, in line with the results discussed in \cite{sannino}. Since it is known that  $R^2$ gravity leads to a scale-invariant spectrum of scalar perturbations, we concluded that small deviations from $f(R)=R^{2}$ are crucial. Such deviations can appear in various form. For instance, one can add other gravitational quadratic terms and the matter fields of the standard model in such a way that the Planck scale is dynamically generated, as in \cite{strumia} and in earlier work reviewed in \cite{adler}. Alternatively, one can add quantum one-loop corrections to $R^{2}$ and resort again to the mechanism advocated by Adler in \cite{adler}.  However, in our approach, the $R^{2}$ term is not interpreted as a loop correction to something else, but as the classical (inflationary) Lagrangian. This choice is also based on theoretical motivations. For example, the addition of the linear Einstein term to $R^{2}$ leads to a tachyonic degree of freedom at tree-level and in flat space \cite{stelle}. This problem persists in the presence of terms like $R_{\mu\nu}^{2}$ which, in addition, lead to multi-field inflation in the Einstein frame description of the theory. The same occurs for terms such as $R\Box^{n}R$ \cite{Wands}. Finally, $R^{2}$ has a de Sitter vacuum solution with arbitrary cosmological constant. 

In principle, the fundamental Lagrangian should also contain the non-minimally coupled Higgs field. On the other hand,  one can assume that, at the onset of inflation, the standard model matter is in the symmetric vacuum, so that the (vacuum) expectation value of the Higgs field is vanishing. Indeed, in de Sitter space  the Higgs potential looks like
\begin{equation}
V=-\frac{m^{2}}{2}{\cal H}^{\dag}{\cal H}+\frac{\lambda}{4}({\cal H}^{\dag}{\cal H})^{2}+\xi R_{\rm ds}{\cal H}^{\dag}{\cal H}\,,
\end{equation}
where $m\sim 125\;\rm{Gev}$ is the standard model Higgs mass and $R_{\rm ds}$ is the de Sitter curvature.  As long as $\xi R_{\rm ds}\geq m^{2}/2$, the symmetry is unbroken and $\braket{{\cal H}}=0$. Then, the term linear in $R$ vanishes, leaving the cause of inflation to  higher-order terms. 

Motivated by these considerations, we study an effective Lagrangians of the form $R^{2}+$ loop corrections. We first show that a finite number of loop corrections leads to spectral indices incompatibles with data. Therefore, we suggest a phenomenological form which mimics a possible resummed Lagrangian for which the values of $r$ and $n_{s}$ are consistent with the Planck data.  In particular, $r$ is significantly larger that the one of the Starobinsky model (SM) but still within the experimental bounds. We will also show that this form of Lagrangian arises naturally in tensor-scalar theories with Coleman-Weinberg quantum-corrected potentials.

Our approach is based on the effective Lagrangian formalism and is driven by phenomenological considerations. As discussed below, the starting point of our work is the calculation of one-loop corrections to quadratic gravity by expanding around a (Euclidean) de Sitter space, as we believe that this is more appropriate for an effective theory of gravity that describes inflation. Instead, the traditional approach to renormalizable theories of gravity, based on the pioneering work of Stelle \cite{stelle}, treats the quantization of fluctuations around flat space (see e.g \cite{QG}). Whether or not these methods are equivalent is an open question.

%%%%%%%%%%%% INTRO  SECTION %%%%%%%%%%%%%%%%%%%%%

\section{The scale invariant $R^2$ model and its deformations}

%%%%%%%%%%%%%%%%%%%%%%%%%%%%%%%%%%%%%%%%

\noindent To begin with, let us recall the most general scale-invariant  Jordan frame action containing the square of the Ricci scalar, the Weyl invariant, and the Higgs doublet ${\cal H}$ non-minimally coupled to gravity:
\bea\label{invlagra}
S_{J}=\int d^{4}x\sqrt{g}\Big[ bR^{2}+aW+\xi R{\cal H}^{2}-(\partial {\cal H})^{2} -{\lambda\over 4} {\cal H}^{4} +\ldots\Big].
\eea
The dots stand for the other quadratic invariants of the metric and  scale invariant operators of the standard model.  Here, the parameters $b$, $\xi$, and $\lambda$ are all dimensionless. This action has been thoroughly investigated in the past, see e.g. \cite{buch0}. Recently, it was reconsidered within a much more large physical context in \cite{strumia}, where it was shown that it leads to an inflationary  model consistent with observations, provided one adds a new scalar field degree of freedom and takes in account the running of  $b$, $\xi$, and $\lambda$. In addition, this model is particularly attractive as it is believed to be renormalizable and asymptotically free \cite{stelle,tomb,fra,avra}, although ghosts are in general present (for a review, see \cite{buch}).

In this paper, we take a different look at this action and we  show that inflation may be realized by  just the first term of eq.\ \eqref{invlagra}, implemented by suitable loop corrections. As explained in the introduction, in the symmetric state, the operators related to the Higgs field vanish until the curvature drops below a certain critical value. In the following, we adopt the point of view that, if by ``quantization'' we understand the process of functional integration over a set of fundamental, non-composite fields, then the effective action has a loop-expansion in the Nambu sense: take a parameter $g$, replace the action $I\to g^{-1}I$ and expand the functional integral in powers of $g$. Then, one sees that a $L$ loop-connected graph has a coefficient $g^{L-1}$. Hence, the one-loop term is given by a suitable ratio of functional determinants. The important point is that we integrate the Euclidean action over the four-sphere, which corresponds to de Sitter space, rather than over flat Euclidean space. In fact, the theory $ R^{2}$ has only a one-parameter family of non-trivial  homogeneous and isotropic solutions, among which we find de Sitter space and the  radiation-dominated Universe. Together with the latter, the other solutions do not seem to have any Euclidean counterpart and presumably lead to an ill-defined functional integral (see \cite{buchdahl} for other remarkable properties of $R^{2}$).

In Ref.\ \cite{trento}, and on the grounds of the results presented in \cite{cognola},  we showed that the one-loop corrections to $R^{2}$ leads to the effective Lagrangian 
\bea\label{fullaction}
f_{\rm eff}(R)=R^{2}\left[1-\gamma\ln\left(R^{2}\over \mu^{2}\right)\right],
\eea
where  $\gamma$ is a small positive parameter and  $\mu$ is a constant that fixes the scale of the corrections (see also \cite{ford1}, and, for the asymptotic safety approach, \cite{alfio}). A similar expression was obtained in \cite{ketov} by a conformal transformation from an Einstein frame action containing a quadratic potential and a cosmological constant. However, as we will show shortly, this Lagrangian inevitably leads to scalar spectrum, which is incompatible with observations so it must be discarded. Motivated by this no-go result, we  softly break the scale invariance with a deformation of the classical $R^2$ Lagrangian that mimics the resummation of higher-loop logarithmic  corrections (in absence of  gravity, see, for example \cite{mk}), namely
\bea\label{resum}
f_{\rm eff}(R)={R^{2}\over\left[1+\gamma\ln\left(R^{2}\over \mu^{2}\right)\right]}\,.
\eea
An example of a theory in which a Lagrangian of this form appears more or less naturally is given by taking Eq.~\eqref{invlagra}, with $b=1$, $\xi=1/6$, and $a=0$, together with the Coleman-Weinberg quantum corrections, which are still valid for these values   \cite{Bytsenko:1994bc}. In the slow-roll approximation we can neglect the kinetic term and the action is extremized at  a value of ${\cal H}$ satisfying the implicit equation
\begin{equation}\label{gianni}
{\cal H}^{2}={R\over 6\lambda}\left[1+\delta+\delta^{{'}}\lambda\log({\cal H}^{2}/\tilde{\mu})\right]^{-1}\,.
\end{equation}
The scale $\tilde{\mu}$, which has the dimension of a squared mass, is necessary to make the logarithm dimensionless and $\delta=O(\hbar)$, $\delta^{'}=O(\hbar)$ are small numbers. The perturbative result is considered non-trustable near ${\cal H}=0$ in flat space, although this is consistent with Eq.~\eqref{gianni},  or for large values of the field of order $\sim\tilde{\mu}\exp(1/\lambda\delta^{'})$, which is huge. Therefore, we reasonably assume that ${\cal H}^{2}$ has a value around the scale $\tilde{\mu}$. In any case, the solution of the above equation can be given in terms of Lambert functions. By retaining only the relevant numerical factor, we find 
\begin{equation}
{\cal H}^{2}\propto {R\over \lambda \delta'}\, W\left[ {R\over \lambda \delta' \tilde \mu}\exp\left({1+\delta\over \lambda\delta'}\right)\right]^{-1}\,,
\end{equation}
where $W[z]$ is the Lambert function of argument $z$, namely the  solution of $z=We^{W}$. Since $\lambda\delta'\ll 1$, we are in the regime of very large arguments for $W(z)$, so the asymptotic expansion near infinity would give
\begin{equation}\label{giannia}
{\cal H}^{2}={R\over 6\lambda}\left[1+\delta+\delta'\log\left({R^{2}\over \mu^{2}}\right)-\lambda\delta'\log\log\left({R^{2}\over\mu^{2}}\right)+\cdots\right]^{-1}\,,
\end{equation}
where $\mu$ is another convenient mass scale and the dots are terms of order $\log^{j}\log R/\log^{k}R$, $1\leq j\leq k$.
Plugging this result into Eq.~\eqref{invlagra}, and omitting the sub-leading log-log terms, together with the kinetic term (in the slow-roll approximation), gives the on-shell Lagrangian whose leading term has the form postulated in Eq.\eqref{resum}  (the $1+\delta$ term can be rescaled to $1$ in the denominator, since the coefficient of $R^{2}$ is arbitrary). The pole probably reflects the Landau pole of the original theory or it is simply the result of the approximation. Note that, without the mass term, we cannot appeal to the symmetry restoration due to de Sitter curvature that we mentioned in the introduction, so ${\cal H}=0$ is a solution for any curvature and the only available argument to discard it would be the invalidity of the potential around the point ${\cal H}=0$, besides the fact that it would give a trivial lagrangian. On the other hand, near the minimum at $R=R_{\rm min}$ of Eq.~\eqref{resum}, where, as we will show below,  there is no inflation, the Lagrangian reduces to a power-law expansion in $R$, profoundly different from the $\log R+\log\log R+\cdots$ expansion discussed here. From this point of view at least, inflation is very sensitive to ultraviolet behavior.\par
In addition, with the reasonable assumption that $\gamma\ll 1$, the curvature around the minimum is so  small that the effects induced by the Higgs and other standard model fields cannot be ignored and a different physics should sets in. In any case, we will show below that inflation occurs entirely at  larger values than $R_{\rm min}$, independently of the value of $\gamma$.   So, we feel authorized in using Eq.~\eqref{resum} during the inflationary phase (some other argument in favor of this choice will be given in the next sections). However, we may anticipate that the surprising feature of Eq.~\eqref{resum} is that it yields an inflationary phase such that the spectral index, its running, and the tensor-to-scalar ratio depend exclusively on the number of e-folds. The only constraint that relates $\gamma$ and $\mu$ comes from  the amplitude of the scalar power spectrum. 
%even if we are not really able to fully justify this choice of the Lagrangian (some other argument in favor of this choice will be given in the next Section) 
%%%%%%%%%%%%%%%%%%%%%%%%%%%%%%%%%%%%%%%%%%

\section{Inflation in  $f(R)$ theories}

%%%%%%%%%%%%%%%%%%%%%%%%%%%%%%%%%%%%%%%%%%

\noindent In order to obtain the inflationary observables, we introduce a simple and transparent formalism that is valid for all $f(R)$ theories.
Let us consider the generic action in Jordan frame (for reviews on $f(R)$ gravity see e.g. \cite{nojiri, fara, defelice})
\bea\label{Jframe}
S_{J}=\int d^{4}x\sqrt{|g|}f(R).
\eea 
Our goal is to express the usual inflationary observables in both Einstein and Jordan frame in a simple and universal form. The only vacuum equation of motion for a homogeneous and isotropic Universe with metric $ds^{2}=-dt^{2}+a^{2}d\vec x^{2}$  is  
\bea\label{eom}
3XH^{2}&=&{1\over 2}(XR-f)-3H\dot X\,,
\eea
where the dot represents a derivative with respect to the (Jordan frame) cosmic time $t$, $H=a^{-1}\dot a$ is the Hubble function,  $R\equiv 6(2H^{2}+\dot H)$, and
 $X\equiv {df(R)/ dR}$.
The conformal transformation $\tilde g_{\mu\nu}=Xg_{\mu\nu}$ brings the action \eqref{Jframe} into the canonical form in Einstein frame 
\bea\label{eframeaction}
S_E=\int d^4x\sqrt{|\tilde g|}\left[{M^2\over 2}\tilde R-{1\over 2}(\tilde\partial\tilde\phi)^2-V(\tilde\phi)\right],
\eea
where
\bea
V(\tilde\phi)={M^2\over 2}\left(XR-f(R)\over X^{2}\right),
\eea
and $X$ and $\tilde \phi$ are related by
\bea\label{xofphi}
\tilde\phi=\sqrt{3\over 2}M\ln( X)\,.
\eea
Usually, the parameter $M$  is identified with the Planck mass $m_{p}$ under the hypothesis that the action \eqref{eframeaction} describes also the low-energy limit of the theory. However, since we will deal with the  scale-invariant  Lagrangian  \eqref{resum}, this identification is not strictly speaking justified. Nevertheless, for now we keep a conservative point of view by setting $M=m_{p}$ and we will comment below on alternative choices.
If $X(R)$ is positive definite and invertible, we can always write a derivative with respect to $\tilde\phi$ in terms of a derivative with respect to $R$. In particular, we can express the slow-roll parameters as (the prime indicates a functional derivative with respect to $R$)
\bea\label{eps}
\epsilon&=&{M^2\over 2}\left(d\ln(V)\over d\tilde\phi\right)^2={(XR-2f)^2\over 3(XR-f)^2},\\\non
\eta&=&{M^2\over V}{d^2V\over d\tilde\phi^2}={2(XR-4f)\over 3(XR-f)}+{2X^2\over 3(XR-f)X'}\,\,,\\\non
\xi^{2}&=&{M^{2}\over V^{2}}{dV\over d\tilde\phi}{d^3V\over d\tilde\phi^3}={4(XR-2f)(X^{3}X''+X'^{3}XR-8X'^{3}f+3X^{2}X'^{2})\over 9X'^{3}(XR-f)^{2}},
\eea
from which we  construct the spectral index, its running, and the tensor-to-scalar ratio defined as
\bea\label{indices}
n_{s}=1-6\epsilon+2\eta, \quad r=16\epsilon, \quad {dn_{s}\over d\ln k}=16\epsilon\eta-24\eta^{2}-2\xi^{2}.
\eea
With the help of the definition \eqref{xofphi}, we can also define the number of e-fold (in Einstein frame) as a function of $\tilde\phi$ or $R$ according to 
\bea\label{Nefold}
\tilde N(\tilde\phi)={1\over M^{2}}\int  V\left(dV\over d\tilde\phi\right)^{-1}d\tilde\phi={3\over 2}\int {V\over V'}{X'^{2}\over X^{2}}dR.
\eea
We stress that these formulae are valid for any $f(R)$ theory and hold whenever $X(R)$ is positive definite and invertible.

With the help of these formulae we can give an additional phenomenological justification of Eq.\ \eqref{resum}. Suppose that the full Lagrangian is a deformation of the SM form induced by $n$ loop corrections, so it can be written as
\bea\label{deformedstaro}
f(R)=\xi {\cal H} R+a_{2}R^{2}g(z),\quad z=\ln \left(R^{2}\over \mu^{2}\right)\,,
\eea
for some constant $a_{2}$ and $\xi$.  Here we stress again that the role of inflaton is played by the gravitational scalaron, and that $\cal H$ may play a role only at the end of  inflation. We can reasonably assume that the function $g(z)$ can be written, at least in the regime of interest, as $g(z)=1+\gamma_{1}z+\gamma_{2}z^{2}+\ldots+\gamma_{n}z^{n}$ with $\gamma_{1}\gg\gamma_{2}\gg\ldots\gg\gamma_{n}$ . By using the first two equations of \eqref{eps} and the expressions of $n_{s}$ and $r$ we find that (for small $r$ and  $\gamma_{1}\ll1$) $r\simeq 3(1-2\gamma_{1})(n_{s}-1)^{2}$. This shows that the corrections to the predictions of the SM (i.e. $r\simeq 3(n_{s}-1)^{2}$) are relatively small but in principle measurable, as already noticed in several papers (see e.g. \cite{sannino}). 

Let us now consider eq.\ \eqref{deformedstaro} with $\xi=0$. In this case,  we find that $n_{s}\simeq 1+r/8$, which is patently in contrast with observations. However, if we assume that all the logarithmic corrections give a contribution of the form \eqref{resum}, then $r\simeq 8(1-n_{s})/3$, which lies within the experimental bounds and represents a class of models disjoint from the SM class. For $n_{s}\simeq 0.968$ \cite{P15}, this relation yields $r=0.085$.

\begin{figure}[h]
 \includegraphics[scale=.5]{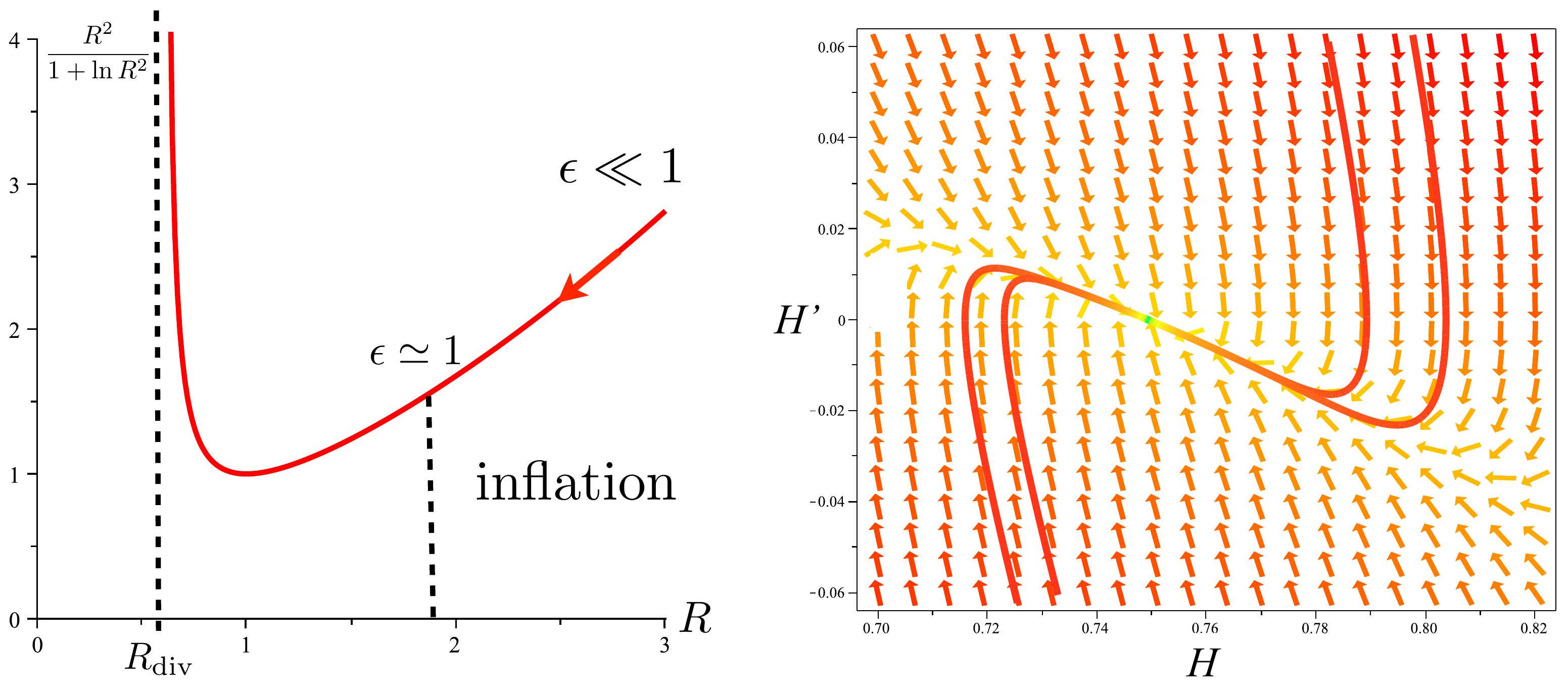}
\caption{On the left we show the qualitative plot of the Lagrangian \eqref{resum} with the flow indicated by an arrow. On the right we show that the point $R=R_{\rm div}$ corresponds to a stable attractor in the plane $(H,H')$. Thus, the system naturally flows from large values of curvature (and small slow-roll parameters) towards the end of inflation and beyond, where other fields are expected to intervene in the dynamics. }
 \label{plot}
 \end{figure}

As discussed above,  eq.\ \eqref{resum} is valid only in the range $R>R_{\rm div}$, where it has a global minimum at $R_{\rm min}=\mu\exp [(\gamma-1)/2\gamma]$ so both $f'(R)$ and $f''(R)$ are positive for $R>R_{\min}$, fulfilling the conditions of \cite{app}. The numerical analysis of the equations of motion reveals that $R_{\rm div}$ is an attractor therefore  the system, during inflation, evolves from a value  $R\gg \mu$ towards $R_{\rm div}$, see Fig.\ (\ref{plot}). The deviation of Eq.\ \eqref{resum} from the purely quadratic Lagrangian is of the order $1-\gamma (R^{2}/\mu^{2})$, thus, for sufficiently small $\gamma$, the solution of the equations of motion is close to a pure de Sitter evolution up to $R$ of any given order of magnitude larger than $\mu$.

%%%%%%%%%%%%%%%%%%%%%%%%%%%%%%%%%%%%%%%%%%

\section{Inflationary observables }

%%%%%%%%%%%%%%%%%%%%%%%%%%%%%%%%%%%%%%%%%%

\noindent To check the consistency of our model,  we must verify that inflation last long enough. With the help of eqs.\ \eqref{eps}, we find that
\bea\label{slowrollZ}
\epsilon&=&{4\gamma^{2}\over 3(1+\gamma z-2\gamma)^{2}},\quad\eta=-{16\gamma^{3}\over 3(1+\gamma z-2\gamma)\left[(1+\gamma z)^{2}-3\gamma(1+\gamma z)+4\gamma^{2} \right]}.
\eea
By combining these expressions with the first two eqs.\ of \eqref{indices} we find that
\bea
n_{s}-1= -{r(\pm 9r+28\sqrt{3r}\pm 96)\over 8(\pm 3r+4\sqrt{3r}\pm 32)}  \simeq- {3r\over 8}\mp {\cal O }(r^{3/2}),
\eea
where the second equality holds for small $r$. This confirms, to leading order, the relation $r\simeq 8(1-n_{s})/3$ found above. It is remarkable that this relation does not depend on $\gamma$ and $\mu$. To fix a point in the $(n_{s},r)$ plane, we must compute the number of e-folds before the end of inflation, which typically occurs when $\epsilon(z,\gamma)= 1$. This equation has two solutions, of which only one is at a value of $R>R_{\rm min}$, corresponding to
\bea
z_{\rm end}=-{1\over  \gamma}+2+{2\sqrt{3}\over 3}.
\eea
We note in passing that this result implies that inflation occurs entirely for $R>R_{\rm min}>R_{\rm div}$, namely not only in the range of validity of our theory but also in the stability range, where $X$ and $dX/dR$ are positive definite  \cite{app}.
By integrating eq.\ \eqref{Nefold}  we find that
\bea
\tilde N(z)={3z^{2}\over 16}-{3z\over 2}+{3z\over 8\gamma}+{3\over 4}\ln\left[(1+\gamma z)^{4}\over (1+\gamma z-\gamma)\right].
\eea
At a given number $\tilde N^{\star}$ of e-folds before the end of inflation, the corresponding value of $z_{\rm ex}$ is then implicitly determined by 
\bea\label{Nex}
\tilde N(z_{{\rm ex}})-\tilde N(z_{\rm end})=\tilde N^{\star}.
\eea
The spectral index, its running, and the tensor-to-scalar ratio are finally obtained numerically by inserting $z_{\rm ex}$ in the expressions \eqref{slowrollZ} and \eqref{indices}. One surprising characteristics is that the results do not depend on $\gamma$ but only on $\tilde N^{\star}$. In table I, we report the numerical values of $n_{s}$, $r$, and $dn_{s}/d\ln k$ for a range of $\tilde N^{\star}$

\begin{table}[ht]
\begin{tabular}{|c||c|c|c|}
\hline
   $\tilde N^{\star}$ & $n_{s}$ & $r$  &  $dn_{s}/d\ln k$  \\
   \hline\hline
  40 & 0.9661 & 0.084  &   -0.0008 \\
  \hline
  45 & 0.9697 & 0.075 & -0.0006 \\
  \hline
  50 & 0.9727 & 0.068 & -0.0005 \\
  \hline
\end{tabular}
\caption{Values of $n_{s}$, its running,  and $r$ corresponding to three values of the number of e-folds before the end of inflation.}
\end{table}

We note that, in order to fit the experimental value $n_{s}=0.968\pm 0.006$ \cite{P15} we need to take a number of e-fold which is lower than the standard interval $50<\tilde N^{\star}<60$. However, it is known that for non-polynomial (in Einstein frame) models of inflation such a range can be different, depending on the details of the reheating mechanism \cite{liddle}. We also note that the tensor-to-scalar ratio is about ten times larger than the one predicted by the SM but still lies within the experimental Planck bound $r\lesssim 0.09$ \footnote{In the SM one has $r=192/(4\tilde N+3)^{2}$, which amounts to $r=0.007$ for $\tilde N=40$.}.  The running of the spectral index is negative, small and fully compatible with the Planck result $dn_{s}/d \ln k=-0.003\pm 0.007$  \cite{P15}. Although $n_{s}$, $dn_{s}/d \ln k$, and $r$ are independent of $\gamma$ and $\mu$, the  amplitude $A_{s}$ of the power spectrum of the curvature perturbations is not. In our model (with the assumption that $M$ is the same as the Planck mass) we find the expression
\bea\label{amplitude}
A_{s}={V\over 24\pi^{2}M^{4}\epsilon}={\mu(1+\gamma z)^{2}(1+\gamma z-2\gamma)^{3}\over 512\,M^{2}\pi^{2}\gamma^{2}(1+\gamma z-\gamma)^{2} },
\eea
which must be evaluated at the horizon exit $z=z_{\rm ex}$. By assuming the typical value $A_{s}\simeq 2\times 10^{-9}$, we find that ${\sqrt{\mu}/ M}\simeq {5\times 10^{-5}/ \sqrt{\gamma}}$. The parameter $\gamma$ is assumed to be a small number, and only when $\gamma\sim 10^{-9}$ the mass scale $\sqrt{\mu}$ approaches the value of the Planck mass $M$. With the help of eqs.\ \eqref{xofphi} and \eqref{Nex}, we can write $\tilde\phi$ at a generic $\tilde N^{\star}$ as
\bea
{\tilde\phi^{\star}\over M}=F(\tilde N^{\star})-{\sqrt{6}\over 2}\left(\ln \gamma+{1\over 2\gamma}\right),
\eea 
where the first term is a complicate algebraic  function of $\tilde N^{\star}$ only. If, for example, we require that the value of $\tilde \phi$ at horizon exit is of the order of $5M$, as in the SM \cite{staro}, we find that, for $\tilde N^{\star}=40$, $\gamma\simeq 0.087$ in line with the requirement that $\gamma\ll1$. In Jordan frame, this value corresponds to $R_{\rm ex}\simeq 3\times 10^{-8} M^{2}$ and to a Hubble parameter that can be estimated to be of the order of $H_{\rm ex}=\sqrt{R_{\rm ex}/12}\simeq \sqrt{2\mu}=5\times 10^{-6} M $, similarly to the SM. This shows that our model can be compared to the SM in terms of energy scales and spectral index. In the recent paper \cite{vernov}, similar results concerning $n_s$ and $r$ have been obtained  without including the Hilbert-Einstein term to the action, but taking into account quantum corrections to the potential, coming from the RG-equations related to scalar electrodynamics.

%%%%%%%%%%%%%%%%%%%%%%%%%%%%%%%%%%%%%%%%%%

\section{Conclusions}

%%%%%%%%%%%%%%%%%%%%%%%%%%%%%%%%%%%%%%%%%%

\noindent In this paper we entertained the idea that the inflationary Universe can be entirely described by a purely quadratic gravitational effective theory, provided loop corrections are taken in account. We find that the spectral index of scalar perturbations matches the Planck data, while the scalar-to-tensor ratio is about ten times larger than the one of the Starobinsky model. Remarkably, these predictions are independent of the parameters of the theory.
It is worth noticing that our model is not, in principle, affected by a transplanckian problem. It is know, from the Lyth bound \cite{lythbound}, that in single-field inflation, in order to have a non-negligible value of $r$, one needs a planckian excursion of the inflaton field, according to $\Delta\phi/m_{p}\sim \int_{0}^{\tilde N^{\star}} d\tilde N\sqrt{r}$. If future data analysis will confirm that $r$ is of the order of $1/10$, it will be difficult to claim that quantum gravitational effects should not be taken in account in these models. In our case, the situation is different as the scale $M$ that appears in \eqref{xofphi} does not necessarily match the Planck mass. Thus, in general, the expression $\Delta\phi/m_{p}$ should contain the ratio $M/m_{p}$ that can relax the Lyth bound. This is consistent with the fact that the classical part of our model is scale-invariant and physical scales appear only after inflation, when standard model particles emerge. 

The details of the exit from inflation in our model are not clear yet, so further investigations are necessary. On one hand, we have shown that  $f(R)$ has a global minimum at $R = R_{\rm min}$ that is similar to the one in SM. However,  since it is not located at $R=f(R)=0$, the standard reheating mechanism via oscillations of the inflaton field does not work here.  On the other hand, we saw in Sec.\ 2,  that our theory is in fact ``dual", to the leading term, to a conformally coupled scalar-tensor theory with Coleman-Weinberg quantum corrections of the potential, which is known to offer a graceful exit, at least in some area of its parameter space.  We also argued that, at the end of inflation, we expect that the Higgs sector becomes relevant, if not dominant, for the post-inflationary evolution therefore the pole in the effective action has no real physical meaning. In addition to this, we also expect that the effects of the Higgs sector are crucial also for the calculation of the exact number of e-folds, and to assess whether our model suffers from some fine-tuning of the parameters.  How exactly the Higgs sector intervenes at the end of inflation is an open and fundamental question that will be hopefully addressed in future work.

\end{document}